\documentclass[aps,pra,twocolumn,showpacs,notitlepage,superscriptaddress,letterpaper,amsmath,amssymb]{revtex4-2}%draft
\usepackage{graphicx}% Include figure files
\usepackage{dcolumn}% Align table columns on decimal point
\usepackage{bm}% bold math
\usepackage{braket}
\usepackage{lipsum}
\usepackage{mathtools}
\usepackage{hyperref}

\begin{document}
%\preprint{APS/123-QED}

%\title{Observation of entanglement between single optical and microwave photons emitted by a chip-scale transducer}
% \title{Observation of entanglement between microwave and optical flying qubits from a single-chip transducer}
% \title{Quantum entanglement between a single optical photon and a single microwave photon}
\title{Quantum entanglement between optical and microwave photonic qubits}
% \title{Entanglement between photonic qubits at optical and microwave frequencies}

\author{Srujan Meesala}
\thanks{These authors contributed equally}
\affiliation{Kavli Nanoscience Institute and Thomas J. Watson, Sr., Laboratory of Applied Physics, California Institute of Technology, Pasadena, California 91125, USA}
\affiliation{Institute for Quantum Information and Matter, California Institute of Technology, Pasadena, California 91125, USA}
\author{David Lake}
\thanks{These authors contributed equally}
\affiliation{Kavli Nanoscience Institute and Thomas J. Watson, Sr., Laboratory of Applied Physics, California Institute of Technology, Pasadena, California 91125, USA}
\affiliation{Institute for Quantum Information and Matter, California Institute of Technology, Pasadena, California 91125, USA}
\author{Steven Wood}
\thanks{These authors contributed equally}
\affiliation{Kavli Nanoscience Institute and Thomas J. Watson, Sr., Laboratory of Applied Physics, California Institute of Technology, Pasadena, California 91125, USA}
\affiliation{Institute for Quantum Information and Matter, California Institute of Technology, Pasadena, California 91125, USA}
\author{Piero Chiappina}
\affiliation{Kavli Nanoscience Institute and Thomas J. Watson, Sr., Laboratory of Applied Physics, California Institute of Technology, Pasadena, California 91125, USA}
\affiliation{Institute for Quantum Information and Matter, California Institute of Technology, Pasadena, California 91125, USA}
\author{Changchun Zhong}
\affiliation{Pritzker School of Molecular Engineering, The University of Chicago, Chicago, IL 60637, USA}
\author{Andrew D. Beyer}
\affiliation{Jet Propulsion Laboratory, California Institute of Technology, 4800 Oak Grove Dr, Pasadena, California 91109, USA}
\author{Matthew D. Shaw}
\affiliation{Jet Propulsion Laboratory, California Institute of Technology, 4800 Oak Grove Dr, Pasadena, California 91109, USA}
\author{Liang Jiang}
\affiliation{Pritzker School of Molecular Engineering, The University of Chicago, Chicago, IL 60637, USA}
\author{Oskar~Painter}
\email{opainter@caltech.edu}
\homepage{http://copilot.caltech.edu}
\affiliation{Kavli Nanoscience Institute and Thomas J. Watson, Sr., Laboratory of Applied Physics, California Institute of Technology, Pasadena, California 91125, USA}
\affiliation{Institute for Quantum Information and Matter, California Institute of Technology, Pasadena, California 91125, USA}
\affiliation{Center for Quantum Computing, Amazon Web Services, Pasadena, California 91125, USA}

\date{\today}
\maketitle

%% abstract
\textbf{Entanglement is an extraordinary feature of quantum mechanics. Sources of entangled optical photons were essential to test the foundations of quantum physics through violations of Bell's inequalities. More recently, entangled many-body states have been realized via strong non-linear interactions in microwave circuits with superconducting qubits. Here we demonstrate a chip-scale source of entangled optical and microwave photonic qubits. Our device platform integrates a piezo-optomechanical transducer with a superconducting resonator which is robust under optical illumination. We drive a photon-pair generation process and employ a dual-rail encoding intrinsic to our system to prepare entangled states of microwave and optical photons. We place a lower bound on the fidelity of the entangled state by measuring microwave and optical photons in two orthogonal bases. This entanglement source can directly interface telecom wavelength time-bin qubits and GHz frequency superconducting qubits, two well-established platforms for quantum communication and computation, respectively.}

%% motivation
Our work arises from the need for scalable techniques to connect superconducting quantum processors of increasing size and complexity \cite{Kjaergaard2020}. In analogy with classical information processing networks, quantum networks have recently been envisioned with optical channels as low loss, room temperature links between superconducting processors cooled in separate dilution refrigerator nodes. Over the past two decades, optical distribution of remote entanglement has been achieved with atoms, ions, color centers, quantum dots and mechanical oscillators \cite{Moehring2007, Weinfurter2012, Bernien2013, Delteil2016, Riedinger2018}. The majority of such demonstrations towards optical quantum networks rely on the DLCZ protocol \cite{Duan2001} and its variants, which are based on a quantum light-matter interface in each node of the network. In these experiments, entanglement is heralded between distant matter qubits through interference of emitted optical photons followed by single photon detection measurements. Remarkably, the fidelity of the entangled pair is insensitive to optical losses since the heralding operation naturally selects events within the computational subspace. Popular variants of this scheme also provide insensitivity to differences in optical path length, and rely on preparation of entangled states between the matter qubit and an optical photonic qubit as a key ingredient \cite{BarretKok2005,Blinov2004,Togan2010,Gao2012}. Recently, there have been proposals to develop this fundamental capability for superconducting qubits by using quantum transducers to prepare microwave-optical Bell states \cite{Zhong2020TimeBin, Zhong2020FreqBin}. In this work, we bring this first enabling step of a well-established optical entanglement distribution toolkit into the realm of superconducting microwave circuits. 

%% background of the field + quick summary of main results
Microwave-optical entanglement generation with transducers has been elusive due to relatively weak nonlinearities, noise from parasitic optical absorption and losses from device integration challenges. In recent progress, a bulk electro-optic transducer was used to infer continuous variable entanglement between microwave and optical fields at the output ports of the device \cite{sahu2023}. In contrast, the entanglement fidelity in our approach is insensitive to losses in collection and detection of optical and microwave photons which are inevitable in a realistic experimental setting. Moreover, our discrete variable approach offers the advantage that a microwave photonic qubit can be directly mapped onto a matter qubit such as a transmon with high fidelity \cite{KurpiersTimeBin2019}. Our demonstration builds on recent technical advances with piezo-optomechanical transducers, where integration with light-robust superconducting circuits has allowed the generation of microwave-optical photon pairs via spontaneous parametric down-conversion (SPDC) \cite{Meesala2023}. We drive emission from such a transducer into dual-rail optical and microwave photonic qubits, each containing exactly one photon in the microwave and optical output ports, respectively. To define the optical photonic qubit, we use a time-bin encoding, the preferred choice in long-distance optical quantum communication \cite{Marcikic2004}. To define the microwave photonic qubit, we use two orthogonal  modes which naturally arise from strongly hybridized acoustic and electrical resonances in our device. We verify entanglement by correlating microwave quantum state tomography results with detection of a single optical photon in a chosen time-bin, or coherent superposition of time-bins achieved with a time-delay interferometer. 

%%%%%%%%%%%%%%%%%%%%%%%%%%%%%%%%%%%%%%%%%%%%%%%%%%%%%%%%%%
% % device intro figure
\begin{figure}[ht]
\includegraphics[width=0.48\textwidth,scale=1]{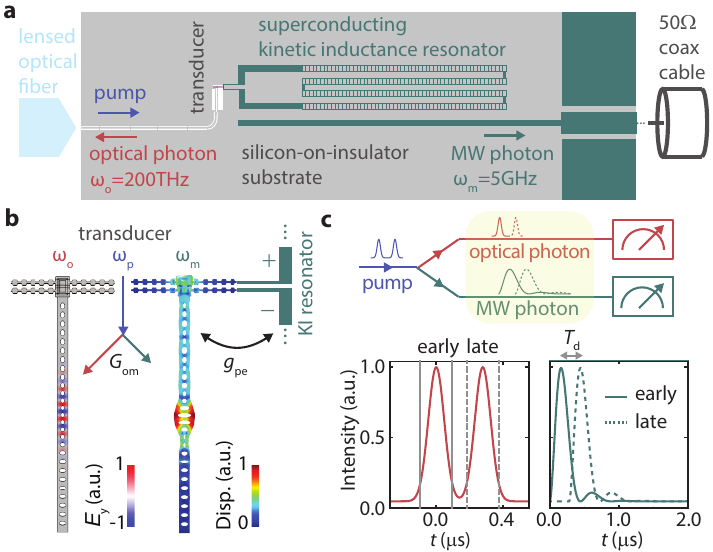}
\caption{\textbf{Microwave-optical entanglement source}. \textbf{a.} Simplified schematic of various components of the chip-scale microwave-optical entanglement source, not shown to scale to aid presentation. The terminals of the superconducting  kinetic inductance resonator are galvanically connected to the electrical terminals of the piezo-optomechanical transducer, shown in detail in panel b. The optical cavity in the transducer is coupled to an optical waveguide, which terminates on the left edge of the chip, where a lensed optical fiber is used to launch pump pulses into the device, and to collect emitted optical photons in the reverse direction. Microwave (MW) photons emitted by the device are capacitively coupled from the superconducting resonator to an on-chip transmission line and eventually collected in a 50ohm coax cable. \textbf{b.} Illustration of the SPDC process in the transducer where a pump photon at frequency, $\omega_{\mathrm{p}}$ decays into optical and microwave excitations at frequencies, $\omega_{\mathrm{o}}$ and $\omega_{\mathrm{m}}$, respectively due to the parametric optomechanical interaction at a rate, $G_{\mathrm{om}}$. The microwave excitation is shared between the transducer acoustic mode and the electrical mode of a superconducting kinetic inductance (KI) resonator, which are strongly hybridized with the piezoelectric interaction strength, $g_{\mathrm{pe}}$. Simulated profiles of the optical electric field (left) and microwave acoustic displacement field (right) in the transducer are shown. \textbf{c.} Schematic for generation of Bell states between dual-rail optical and MW photonic qubits. Two consecutive Gaussian pump pulses induce emission of single optical and MW photons into early and late modes centered at their respective frequencies. Theoretically calculated intensity envelopes of these modes are shown when the device is excited with two Gaussian pump pulses separated by a time delay, $T_\text{d}$~=~279ns, which ensures orthogonality of early and late modes used for the dual-rail encoding.}
% \vspace{-5mm}
\label{fig1}
\end{figure}
%%%%%%%%%%%%%%%%%%%%%%%%%%%%%%%%%%%%%%%%%%%%%%%%%%%%%%%%%%

%% device description
Fig.\,\ref{fig1}a shows a simplified schematic of our microwave-optical entanglement source, which we operate in a dilution refrigerator setup at a temperature of $\sim$20mK. Details of the device geometry and fabrication process have been provided in previous work \cite{Meesala2023}. Pump laser pulses are used to excite a piezo-optomechanical transducer containing a silicon optomechanical crystal resonator which supports optical and acoustic resonances at frequencies, $\omega_{\mathrm{o}} \approx 2\pi\times\,200$THz and $\omega_{\mathrm{m}} \approx 2\pi\times\,5$GHz, respectively. Co-localized telecom photons and microwave phonons in this wavelength-scale resonator can interact via radiation pressure and the photoelastic effect \cite{SMeenehan}. By tuning the pump laser frequency to $\omega_\text{p}=\omega_\text{o}+\omega_\text{m}$ we induce spontaneous parametric down-conversion into a photon-phonon pair at frequencies $\omega_\text{o}$ and $\omega_\text{m}$, as illustrated in Fig.\,\ref{fig1}b. The optomechanical interaction occurs with a parametrically enhanced strength, $G_{\mathrm{om}} = \sqrt{n_{\mathrm{p}}}g_{\mathrm{om}}$, where $n_{\mathrm{p}}$ is the number of intra-cavity pump photons and $g_{\mathrm{om}}/2\pi = 270$kHz is the optomechanical coupling rate at the single photon and phonon level in the device under study. Single phonons from the SPDC process are converted into single microwave photons in a niobium nitride superconducting kinetic inductance resonator tuned into resonance with the transducer acoustic mode. This conversion process is mediated by a compact aluminum nitride piezoelectric component in the transducer, and occurs at a rate, $g_{\mathrm{pe}}/2\pi = 1.2$MHz, which exceeds damping rates of the acoustic and electrical modes in the device under study \cite{SI}. Finally, as shown in Fig.\,\ref{fig1}a, both optical and microwave photons emitted from the device decay into on-chip waveguides and are routed into a lensed optical fiber and a 50ohm microwave coaxial cable, respectively.

%% Bell state generation concept 
To prepare entangled microwave-optical states, we excite the device with two consecutive pump pulses. Each pump pulse can produce a microwave-optical photon-pair in well-defined temporal modes, separated in time and centered at frequencies, $\omega_{\mathrm{m}}$ and $\omega_{\mathrm{o}}$ in the microwave and optical outputs, respectively. Fig.\,\ref{fig1}c shows theoretically expected intensity envelopes of these `early' and `late' modes in the optical and microwave output ports of the device. The optical early and late mode envelopes adiabatically follow the pump pulses since the optomechanical interaction strength, $G_{\mathrm{om}}$ is much smaller than the decay rate of the optical cavity. In our experiments, we use Gaussian pump pulses with two sigma width, $T_{\mathrm{p}}=96$ns, and define time-bin modes for dual-rail encoding of the optical photonic qubit. The corresponding microwave early and late modes exhibit damped oscillatory behavior due to strongly coupled electrical and acoustic resonators in the device. By matching the duration between the pump pulses, $T_\text{d}=279$ns to the electro-acoustic oscillation period expected from microwave spectroscopy \cite{SI}, we can achieve orthogonal early and late modes for dual-rail encoding of the microwave photonic qubit. Physically, a single phonon scattered by the SPDC process into the electro-acoustic coupled mode system oscillates between the acoustic and electrical resonators as it preferentially decays into the microwave output waveguide. With our specific choice of the pump pulse delay, $T_\text{d}$, we excite the transducer with the late pump pulse precisely at the moment when the phonon from the early pump pulse has been maximally swapped into the electrical mode. In this setting, the joint wavefunction of early and late modes in the optical and microwave output ports of the device can be written as
\begin{align}\label{eq:TimeBinWave}
\ket{\Psi}  \approx \,& \ket{00}_\text{o}\!\ket{00}_\text{m} \nonumber\\
&+\sqrt{p}\left(\ket{10}_\text{o}\!\ket{10}_\text{m}+e^{i\phi_{\mathrm{p}}}\ket{01}_\text{o}\!\ket{01}_\text{m}\right)
+\mathcal{O}(p),
\end{align}
\noindent where $\ket{kl}_\text{o}$ ($\ket{kl}_\text{m}$) denotes the direct product of a $k$-photon state in the early mode and an $l$-photon state in the late mode on the optical (microwave) output port. The phase $\phi_{\mathrm{p}}$, is set by the relative phase between early and late pump pulses. When the scattering probability, $p\ll1$, the $\mathcal{O}(p)$ terms may be neglected and detection of a single optical photon can be used to post-select an entangled state between an optical photonic qubit in the $\{\ket{10}_\text{o}, \ket{01}_\text{o}\}$ manifold and a microwave photonic qubit in the $\{\ket{10}_\text{m}, \ket{01}_\text{m}\}$ manifold. The state vectors within curly brackets define the native measurement basis of the photonic qubits, which we call the Z-basis in reference to the north and south poles of the Bloch sphere. We refer to a rotated measurement basis on the equator of this Bloch sphere as the X-basis. To verify entanglement, we characterize correlations between the photonic qubits in these two orthogonal bases. In our experiments, we operate with optical pulses with a peak power of 83nW corresponding to an intra-cavity optical photon number, $n_\text{p}=0.8$, which leads to $p=1.0\times10^{-4}$. The power level is chosen to limit noise added due to pump-induced heating of the transducer, and thereby, preserve microwave-optical entanglement. With external optical collection efficiency, $\eta_{\mathrm{opt}}=5.5\times10^{-2}$ and a pulse repetition rate, $R=$~50kHz, we detect heralding events at a rate, $R_{\text{click}}=0.26\mathrm{s}^{-1}$. Microwave phonon to photon conversion is expected to occur with efficiency, $\eta_{\text{mw}}=0.59$ based on the piezoelectric coupling rate and microwave damping rates in our device.

%%%%%%%%%%%%%%%%%%%%%%%%%%%%%%%%%%%%%%%%%%%%%%%%%%%%%%%%%%
% % Z-basis intensiy correlations figure
\begin{figure}[ht]
\includegraphics[width=0.48\textwidth]{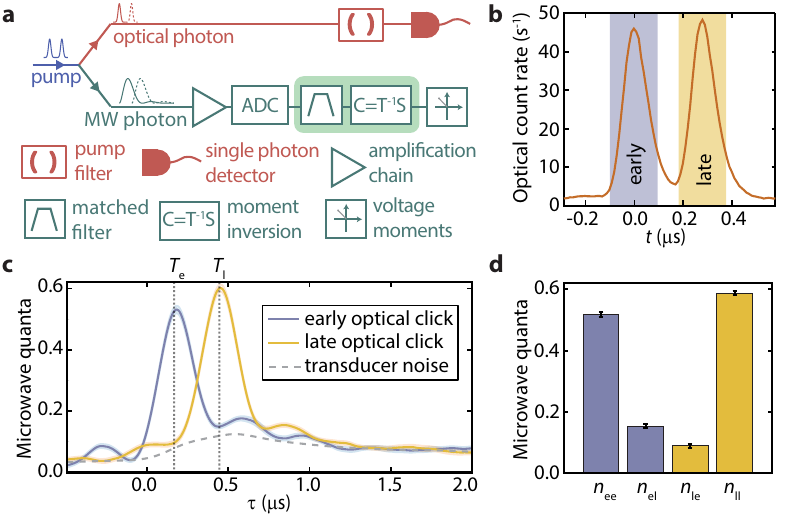}
\caption{\textbf{Z-basis intensity correlations}. \textbf{a.} Simplified schematic of experimental setup used to detect correlations between microwave (MW) and optical emission in the Z-basis of the dual-rail qubits. Shaded green box on the MW detection path indicates post-processing on voltage traces from the heterodyne setup after analog-to-digital conversion (ADC). \textbf{b.} Histogram of single optical photon detection times plotted as a time trace of optical count rate. Shaded vertical regions indicate gating windows used to define early and late optical time-bin modes, $\hat{A}_\text{e}$ and $\hat{A}_\text{l}$. \textbf{c.} Quanta in the transducer microwave output mode defined by a filter function matched to the theoretically expected emission envelope, $f(t-\tau)$ centered at the MW resonance frequency, $\omega_{\mathrm{m}}$ \cite{SI}. The variable readout delay, $\tau$ is shown on the x-axis and the occupation of the mode for a given $\tau$ is shown on the y-axis. Blue and yellow traces show MW output quanta conditioned on early and late optical clicks, respectively, and the dashed gray trace shows unconditional MW output quanta, which correspond to transducer-added noise. Dotted vertical lines indicate readout delays, $T_\text{e}$ and $T_\text{l}$ used to define early and late MW modes, $\hat{C}_\text{e}$ and $\hat{C}_\text{l}$. The conditional traces are an average over $\approx 3\times 10^5$ heralding events. Shaded regions around traces span a confidence interval of two standard deviations about the mean. \textbf{d.} Output quanta, $n_{ij}$ in MW mode $j$ conditioned on an optical click in mode $i$, where $i,j$ run over the early and late modes denoted by $\text{e,l}$.  Data in this panel corresponds to $\approx 3\times 10^5$ heralding events. Error bars indicate $+/-$ one standard deviation.
}
\vspace{-4mm}
\label{fig2}
\end{figure}
%%%%%%%%%%%%%%%%%%%%%%%%%%%%%%%%%%%%%%%%%%%%%%%%%%%%%%%%%%

%Time trace of MW output quanta conditioned on early optical click (blue trace), MW output quanta conditioned on late optical click (yellow trace), and pump-induced noise from the unconditional MW output quanta (dashed gray trace). 

%% Z-basis intensity correlations
We first measure the time-resolved microwave output intensity from the device conditioned on the detection time of single optical photons. This allows us to characterize microwave-optical intensity correlations in the Z-basis. In this measurement, depicted schematically in Fig.\,\ref{fig2}a, optical emission from the device is detected on a superconducting nanowire single photon detector (SNSPD) after transmission through a filter setup to reject pump photons. Fig.\ref{fig2}b shows a histogram of single optical photon detection times revealing two nearly Gaussian envelopes associated with the SPDC signal. We use gating windows of width, $2T_{\mathrm{p}} = 192$ns centered around each pulse to define the early and late optical modes, $\hat{A}_\text{e}$ and $\hat{A}_\text{l}$, respectively. Simultaneously, as shown in Fig.\,\ref{fig2}a, microwave emission from the device is directed to an amplification chain with a near-quantum-limited Josephson traveling wave parametric amplifier (TWPA) as the first stage. The amplified microwave signal is down-converted in a room-temperature heterodyne receiver and the resulting voltage quadratures are sent to an analog-to-digital conversion (ADC) card, allowing us to record a digitized, complex-valued voltage trace for each experimental trial. We capture emission at the microwave resonance frequency, $\omega_{\mathrm{m}}$ using a digital filter matched to the theoretically expected microwave emission envelope, $f(t-\tau)$, where $\tau$ is a variable readout delay \cite{SI}. We then subtract independently calibrated amplifier-added thermal noise of approximately 2.6 quanta via a moment inversion procedure \cite{Eichler2011}. Upon post-selecting measurement records from trials which produced optical clicks, we observe that the microwave intensity conditioned on a late click is delayed with respect to that conditioned on an early click as shown by the solid traces in Fig.\,\ref{fig2}c. These conditional signals contain a finite amount of pump-induced thermal noise from the transducer. Since $p\ll1$, such noise is simply given by the unconditional microwave output intensity, shown with the dashed time trace in Fig.\,\ref{fig2}c. The ratio of the conditional and unconditional microwave intensities yields the normalized microwave-optical cross-correlation function, $g^{(2)}_{AC}$ which reaches a maximum value of 6.8 for early optical clicks, and 5.0 for late optical clicks. Both values exceed the Cauchy-Schwarz bound of 2 for thermal states, and signify non-classical microwave-optical correlations \cite{Meesala2023}. By performing the matched filter operation at optimal microwave readout delays, $T_\text{e}$ and $T_\text{l} = T_\text{e} + T_\text{d}$, shown by the dotted vertical lines in Fig.\,\ref{fig2}c, which maximize the cross-correlation, we define microwave early and late modes, $\hat{C}_\text{e}$ and $\hat{C}_\text{l}$. Fig.\,\ref{fig2}d shows conditional occupations of these modes with the symbol $n_{ij}$ for the occupation of microwave mode $j$ conditioned on an optical click detected in mode $i$. Using these four conditional microwave mode occupations, we define the Z-basis visibility, $V_\text{z}=({n_\text{ee}-n_\text{el}-n_\text{le}+n_\text{ll}})/({n_\text{ee}+n_\text{el}+n_\text{le}+n_\text{ll}})$. For a Bell state without additional noise or microwave loss, we expect $n_\text{ee} = n_\text{ll} = 1$ and $n_\text{el} = n_\text{le} = 0$, resulting in $V_\text{z} = 1$. On the contrary, when the microwave and optical intensities are fully uncorrelated, we expect $V_\text{z} = 0$. From the data in Fig.\,\ref{fig2}d, we find $V_\text{z} = 0.633 \pm 0.014$, indicating significant intensity correlations between early and late modes in the microwave and optical outputs.

%%%%%%%%%%%%%%%%%%%%%%%%%%%%%%%%%%%%%%%%%%%%%%%%%%%%%%%%%%
% % X-basis intensity correlations figure
\begin{figure}[ht]
\includegraphics[width=0.48\textwidth]{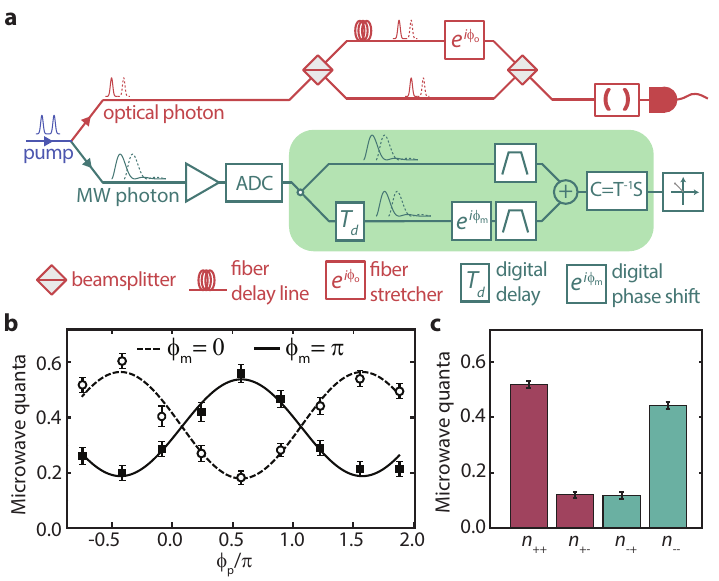}
\caption{\textbf{X-basis intensity correlations}. \textbf{a.} Simplified schematic of experimental setup used to detect correlations between microwave (MW) and optical emission in the X-basis of the dual-rail qubits. A time-delay interferometer in the optical path is used to interfere early and late optical time-bins with a relative phase, $\phi_{\mathrm{o}}$. The interference operation in MW detection with a relative phase, $\phi_{\mathrm{m}}$ is performed in digital post-processing as shown in the shaded green box. \textbf{b.} Output quanta in the MW mode, $(\hat{C}_{e}+e^{i\phi_{\mathrm{m}}}\hat{C}_{l})/\sqrt{2}$, for two phase settings, $\phi_{\mathrm{m}}=0$ (open circles) and $\phi_{\mathrm{m}}=\pi$ (filled squares), conditioned on an optical click in the mode, $(\hat{A}_{e}+e^{i({\phi_{\mathrm{p}}+\phi_{\mathrm{o}}})}\hat{A}_{l})/\sqrt{2}$ at the output of the time-delay interferometer. The relative phase between the pump pulses, $\phi_{\mathrm{p}}$ is varied along the horizontal axis while $\phi_{\mathrm{o}}$ is kept constant at $0.31\pi$. The uncertainty in the calibrated optical phase over the duration of the measurement is $\pm 0.03\pi$. Solid and dashed lines are cosine fits. Data in this plot is acquired at three times the pump power used for the main dataset, and represents an average over $\approx 1\times10^4$ heralding events per optical phase setting. All error bars indicate $+/-$ one standard deviation. \textbf{c.} Output quanta, $n_{ij}$ in MW mode $j$ conditioned on an optical click in mode $i$, where $i,j$ run over the X-basis MW and optical measurement modes denoted by $+,-$ and corresponding to phase settings, $\phi_\text{m} = 0.56\pi, 1.56\pi$ and $\phi_\text{p}+\phi_\text{o} = 0.56\pi, 1.56\pi$, respectively, where $\phi_{\mathrm{o}}$ is kept constant at $0.28\pi$. The uncertainty in the calibrated optical phase over the duration of the measurement is $\pm 0.04\pi$. Data in this panel corresponds to an average over $\approx 7\times10^4$ heralding events for $+$ and $-$ optical outcomes. All error bars indicate $+/-$ one standard deviation.
}
\vspace{-2mm}
\label{fig3}
\end{figure}
%%%%%%%%%%%%%%%%%%%%%%%%%%%%%%%%%%%%%%%%%%%%%%%%%%%%%%%%%%

%% X-basis intensity correlations
This observation of Z-basis correlations is also compatible with a statistical mixture of early and late microwave-optical photon pairs. To rule out this scenario, we characterize intensity correlations in the X-basis, which are indicative of the phase coherence of the entangled microwave-optical state. On the optics side, the measurement basis rotation is performed with a time-delay interferometer inserted into the detection path as shown in Fig.\ref{fig3}\,a. The interferometer is built with a fiber delay line in one arm to achieve the time delay, $T_\text{d}$ = 279ns required to interfere early and late optical time-bins. Additionally, the setup imprints a relative phase between the time-bins, $\phi_\text{o}$, which is actively stabilized through feedback on a piezoelectric fiber stretcher in one of the arms. The relative phase between the pump pulses used to excite the transducer, $\phi_\text{p}$, is controlled using an electro-optic phase modulator \cite{zivari2022chip}. With one output port of the interferometer connected to the single photon detection path, clicks registered on the SNSPD correspond to measurement of a single photon in the mode, $(\hat{A}_\text{e} + e^{i(\phi_\text{p} + \phi_\text{o})}\hat{A}_\text{l})/\sqrt{2}$. The measurement phase, $\phi_\text{p} + \phi_\text{o}$ can be independently calibrated by transmitting coherent optical pulses through the interferometer \cite{SI}. To perform a basis rotation on the microwave side, we add the early and late complex voltage quadratures with a relative phase, $\phi_\text{m}$ in post-processing as shown in Fig.\ref{fig3}\,a. After subtracting amplifier-added noise in a manner similar to the Z-basis measurement, we measure the moments of the microwave mode, $(\hat{C}_\text{e}+ e^{i\phi_\text{m}}\hat{C}_\text{l})/\sqrt{2}$, averaged over experimental trials. In Fig.\ref{fig3}\,b, we show  conditional microwave output quanta measured in the X-basis by fixing $\phi_\text{m} = 0, \pi$. As the optical phase is swept by tuning $\phi_\text{p}$ between the pump pulses, we observe correlation fringes in the conditional microwave intensity, a clear signature of coherence of the entangled microwave-optical state.  Due to the need for data at multiple optical phase settings for this measurement, we used three times higher pump power ($n_\text{p} = 2.4$) compared to the main dataset to speed up acquisition. We then lower the pump power back to $n_\text{p} = 0.8$, the setting used in the Z-basis measurements, and repeat the X-basis measurement for two optical phase settings, $\phi_\text{p}+\phi_\text{o} = 0.56\pi, 1.56\pi$ at which we define the optical modes, $\hat{A}_\pm$, respectively. We measure conditional microwave output quanta in the modes, $\hat{C}_\pm$ for $\phi_\text{m} = 0.56\pi, 1.56\pi$, respectively and obtain the results shown in Fig.\,\ref{fig3}c for the four combinations of the X-basis modes. In a manner similar to the Z-basis correlation measurement, we define the X-basis visibility, $V_\text{x}$, expected to equal 1 for Bell states, and 0 for an equal statistical mixture of early and late microwave-optical photon pairs. With the data in Fig.\,\ref{fig3}c, we observe $V_\text{x} = 0.611 \pm 0.034$. As an additional consistency check, we swept the microwave readout phase, $\phi_\text{m}$ in post-processing, and found that the maximum in $V_\text{x}$ occurs for the phase settings, $\phi_\text{m} = 0.62\pi, 1.62\pi$, offset by $0.06\pi$ from the theoretically expected modes, $\hat{C}_\pm$. This can be attributed to a systematic offset in the calibrated optical phase arising from small differences in optical frequency and polarization between calibration and data acquisition \cite{SI}.   

%Sweeping the microwave readout phase, $\phi_\text{m}$ in post-processing, we find the exact maximum in $V_\text{x}$ occurs for the phase settings, $\phi_\text{m} = 0.62\pi, 1.62\pi$ which is offset by $0.06\pi$ from the theoretically expected modes, $\hat{C}_\pm$ chosen in our analysis. This can arise from small differences in optical frequency and polarization during calibration and data acquisition.  

%which are at a small technical offset of $0.06\pi$ from the theoretically expected modes, $\hat{C}_\pm$ chosen in our analysis. 

%%%%%%%%%%%%%%%%%%%%%%%%%%%%%%%%%%%%%%%%%%%%%%%%%%%%%%%%%%
% % quantum state tomography and conditional probabilities
\begin{figure}[ht]
\includegraphics[width=0.48\textwidth]{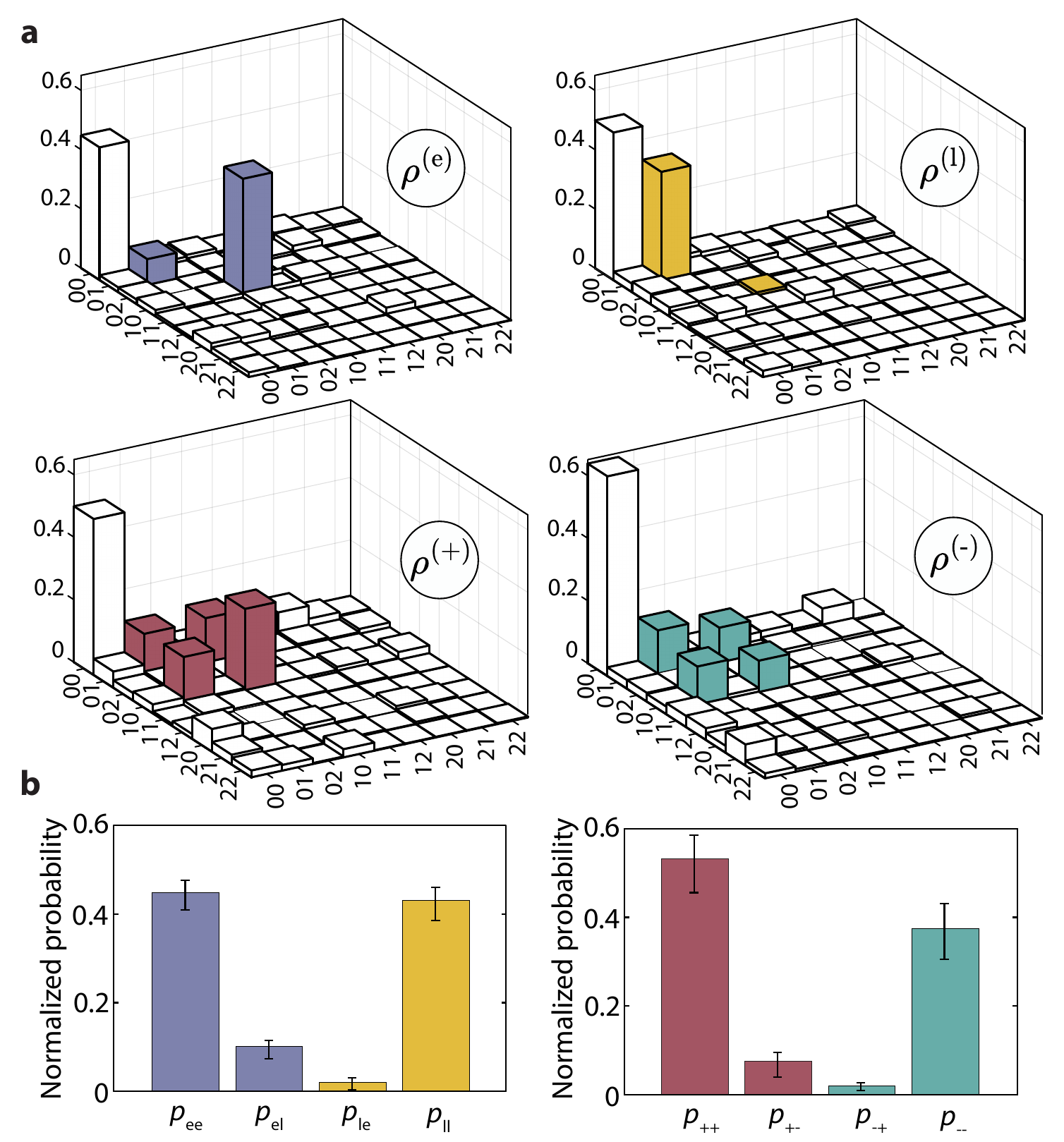}
\caption{\textbf{Quantum state tomography of conditional microwave states}. \textbf{a.} Conditional density matrices, $\rho^{\text{(e)}}, \rho^{\text{(l)}}$, $\rho^{{(+)}}, \rho^{{(-)}}$ of the microwave output state corresponding to early, late, $+$ and $-$ optical clicks, respectively, plotted in the joint Fock basis of early and late microwave modes. The matrices are obtained from a maximum likelihood reconstruction procedure performed over a joint Fock space of up to six photons in each mode, but are plotted in a truncated space of up to two photons in each mode for better visualization. Entries which are expected to be non-zero for a pure microwave-optical Bell state are highlighted in color. \textbf{b.} Conditional probability, $p_{ij}$ of a single photon in microwave mode $j$ conditioned on receipt of an optical click in mode $i$, calculated from the density matrices in panel a after post-selecting the single photon subspace. Error bars denote uncertainties of +/- one standard deviation.
}
\vspace{-2mm}
\label{fig4}
\end{figure}
%%%%%%%%%%%%%%%%%%%%%%%%%%%%%%%%%%%%%%%%%%%%%%%%%%%%%%%%%%

% Quantum state tomography and Bell state fidelity lower bound
The microwave and optical emission from the device exhibit intensity correlations in both Z- and X-bases which are characteristic of a Bell pair prepared from the pure state in Eq.\,\ref{eq:TimeBinWave} via optical detection. However, since the experimentally prepared states have finite transducer-added noise and microwave loss, the conditional microwave intensities have contributions from outside the computational subspace where the dual-rail photonic qubits are defined. In order to characterize entanglement more precisely, we must measure both optical and microwave outputs in the single-photon subspace. Since we operate in an experimental regime where the scattering probability, $p \ll 1$, we fulfill the condition that nearly all optical detection events arise from within the single optical photon subspace \cite{SI}. On the microwave side, we use statistical moments of conditional heterodyne voltages to perform maximum likelihood state tomography in the joint Fock basis of the early and late modes \cite{Eichler2011, Ferreira2022}, and project onto the single photon subspace. This post-selection operation is strictly local and cannot generate microwave-optical entanglement. It can be implemented in practice by performing a parity check \cite{Zhong2020TimeBin}, a well-established capability in circuit quantum electrodynamics. Fig.\,\ref{fig4}a shows conditional density matrices, $\rho^{\text{(e)}}, \rho^{\text{(l)}}, \rho^{{(+)}}, \rho^{{(-)}}$ of the microwave output state corresponding to an optical click in the early, late, $+$ and $-$ modes, respectively. For a pure Bell state, the entries highlighted with colored bars in $\rho^{\text{(e)}},\rho^{\text{(l)}}$ are expected to have the values $\rho^{\text{(e)}}_{10,10} = \rho^{\text{(l)}}_{01,01}=1$, $\rho^{\text{(e)}}_{01,01} = \rho^{\text{(l)}}_{10,10}=0$; likewise, the the entries highlighted with colored bars in $\rho^{\text{(+)}},\rho^{\text{(-)}}$ are expected to have a magnitude of 0.5. The main deviation in the measured conditional microwave states is due to the non-zero vacuum component, primarily from finite conversion efficiency, $\eta_{\text{mw}}=0.59$ of a single phonon into the microwave waveguide. Using the entries highlighted with colored bars in Fig.\,\ref{fig4}a, which denote the computational subspace of the dual-rail photonic qubit, we obtain the conditional probability, $p_{ij}$ of a single microwave photon in mode $j$ conditioned on receipt of an optical click in mode $i$ \cite{SI}. Here $i,j$ run over early and late ($+$ and $-$) modes for Z- (X-) basis measurements, and the results are shown in Fig.\,\ref{fig4}b. Error bars on the probabilities account for statistical error obtained by bootstrapping with replacement over the microwave dataset. These conditional probabilities allow us to establish a lower bound on the Bell state fidelity given by \cite{Zhong2020TimeBin, Blinov2004}
\begin{align}
    F_\text{lb} =&\, \frac{1}{2} (p_{\text{ee}}+p_{\text{ll}}-p_{\text{el}}-p_{\text{le}}\nonumber\\
    & +p_{{++}}+p_{{--}}-2\sqrt{p_{{+-}}p_{{-+}}}).
\end{align}
We find that $F_\text{lb} = 0.794^{+0.048}_{-0.071}$, which exceeds the classical limit of 0.5 by over four standard deviations, indicating the preparation of entangled microwave-optical states. A simple model accounting for pump-induced thermal noise in the transducer \cite{SI} predicts Bell state fidelity exceeding 0.83, which agrees with the measured lower bound. In addition to the primary reduction in the fidelity due to pump-induced noise, we expect smaller contributions due to dark counts and imperfections in optical time-bin interference and microwave mode matching. 

% Conclusion
We envision that the microwave photonic qubit emitted by our entanglement source can be absorbed in a dual-rail superconducting qubit \cite{Zhong2020TimeBin} or a superconducting qutrit \cite{KurpiersTimeBin2019}, both of which can be realized with standard transmons and can enable detection of microwave photon loss errors. Such a transmon module can be connected to the transducer with minimal impact of optical pump light on qubit coherence \cite{JILA_Nondestructive, Arnold2023all-optical, warner2023coherent, vanthiel2023highfidelity}. %This setup would prepare entangled states between the transmon module and a telecom wavelength time-bin qubit which can be verified on the hour timescale given the heralding rate of our transducer.
In the near term, piezo-optomechanical transducers with greater acoustic participation in silicon \cite{Chiappina:23, Zhao:23} and improved thermalization with the substrate \cite{2DOMC} can improve transducer noise performance, enabling microwave-optical entanglement generation rates in the kHz range. Transducer operation in this performance regime can facilitate the integration of superconducting qubit nodes into optical quantum networks for applications in secure communication \cite{Bhaskar2020, Chen2021, Pompili2022} and distributed sensing \cite{Gottesman2012, Khabiboulline2019, Nichol2022}.

% and improved thermalization with the substrate \cite{2DOMC}

% % for arxiv submission only
% \begin{acknowledgments}
% The authors thank A. Butler, G. Kim, M. Mirhosseini and A. Sipahigil for helpful discussions, and B. Baker and M. McCoy for experimental support. We appreciate MIT Lincoln Laboratories for providing the traveling-wave parametric amplifier used in the microwave readout chain in our experimental setup. NbN deposition during the fabrication process was performed at the Jet Propulsion Laboratory. This work was supported by the ARO/LPS Cross Quantum Technology Systems program (grant W911NF-18-1-0103), the U.S. Department of Energy Office of Science National Quantum Information Science Research Centers (Q-NEXT, award DE-AC02-06CH11357), the Institute for Quantum Information and Matter (IQIM), an NSF Physics Frontiers Center (grant PHY-1125565) with support from the Gordon and Betty Moore Foundation, the Kavli Nanoscience Institute at Caltech, and the AWS Center for Quantum Computing. L.J. acknowledges support from the AFRL (FA8649-21-P-0781), NSF (ERC-1941583, OMA-2137642), and the Packard Foundation (2020-71479). S.M. acknowledges support from the IQIM Postdoctoral Fellowship. 
% \end{acknowledgments}
% for journal submission

\bibliography{refs}

\begin{acknowledgments}
The authors thank A. Butler, G. Kim, M. Mirhosseini and A. Sipahigil for helpful discussions, and B. Baker and M. McCoy for experimental support. We appreciate MIT Lincoln Laboratories for providing the traveling-wave parametric amplifier used in the microwave readout chain in our experimental setup. NbN deposition during the fabrication process was performed at the Jet Propulsion Laboratory. This work was supported by the ARO/LPS Cross Quantum Technology Systems program (grant W911NF-18-1-0103), the U.S. Department of Energy Office of Science National Quantum Information Science Research Centers (Q-NEXT, award DE-AC02-06CH11357), the Institute for Quantum Information and Matter (IQIM), an NSF Physics Frontiers Center (grant PHY-1125565) with support from the Gordon and Betty Moore Foundation, the Kavli Nanoscience Institute at Caltech, and the AWS Center for Quantum Computing. L.J. acknowledges support from the AFRL (FA8649-21-P-0781), NSF (ERC-1941583, OMA-2137642), and the Packard Foundation (2020-71479). S.M. acknowledges support from the IQIM Postdoctoral Fellowship. 
 \\
\end{acknowledgments}

\end{document}

\title{Supplementary Information for ``Quantum entanglement between optical and microwave photonic qubits"}

\author{Srujan Meesala}
\thanks{These authors contributed equally}
\affiliation{Kavli Nanoscience Institute and Thomas J. Watson, Sr., Laboratory of Applied Physics, California Institute of Technology, Pasadena, California 91125, USA}
\affiliation{Institute for Quantum Information and Matter, California Institute of Technology, Pasadena, California 91125, USA}
\author{David Lake}
\thanks{These authors contributed equally}
\affiliation{Kavli Nanoscience Institute and Thomas J. Watson, Sr., Laboratory of Applied Physics, California Institute of Technology, Pasadena, California 91125, USA}
\affiliation{Institute for Quantum Information and Matter, California Institute of Technology, Pasadena, California 91125, USA}
\author{Steven Wood}
\thanks{These authors contributed equally}
\affiliation{Kavli Nanoscience Institute and Thomas J. Watson, Sr., Laboratory of Applied Physics, California Institute of Technology, Pasadena, California 91125, USA}
\affiliation{Institute for Quantum Information and Matter, California Institute of Technology, Pasadena, California 91125, USA}
\author{Piero Chiappina}
\affiliation{Kavli Nanoscience Institute and Thomas J. Watson, Sr., Laboratory of Applied Physics, California Institute of Technology, Pasadena, California 91125, USA}
\affiliation{Institute for Quantum Information and Matter, California Institute of Technology, Pasadena, California 91125, USA}

\author{Changchun Zhong}
\affiliation{Pritzker School of Molecular Engineering, The University of Chicago, Chicago, IL 60637, USA}

\author{Andrew D. Beyer}
\affiliation{Jet Propulsion Laboratory, California Institute of Technology, 4800 Oak Grove Dr, Pasadena, California 91109, USA}

\author{Matthew D. Shaw}
\affiliation{Jet Propulsion Laboratory, California Institute of Technology, 4800 Oak Grove Dr, Pasadena, California 91109, USA}

\author{Liang Jiang}
\affiliation{Pritzker School of Molecular Engineering, The University of Chicago, Chicago, IL 60637, USA}

\author{Oskar~Painter}
\email{opainter@caltech.edu}
\homepage{http://copilot.caltech.edu}
\affiliation{Kavli Nanoscience Institute and Thomas J. Watson, Sr., Laboratory of Applied Physics, California Institute of Technology, Pasadena, California 91125, USA}
\affiliation{Institute for Quantum Information and Matter, California Institute of Technology, Pasadena, California 91125, USA}
\affiliation{Center for Quantum Computing, Amazon Web Services, Pasadena, California 91125, USA}

\date{\today}
\maketitle

\section{Device and experiment parameters}

\begin{table}[!ht]
    \centering
    \begin{tabular}{|c|c|c|}
    \hline
    Symbol & Description   &  Value \\
    \hline
     $\omega_\text{o}/2\pi$ & optical mode frequency & 192.02 THz \\     
    \hline
     $\omega_\text{m}/2\pi$ & microwave mode frequency & 5.004 GHz \\     
    \hline
    $g_{\mathrm{om}}/2\pi$ & optomechanical coupling rate & 270 kHz \\     
    \hline
    $g_{\mathrm{pe}}/2\pi$ & piezoelectric coupling rate & 1.2 MHz \\     
    \hline
    $\kappa_\text{e,o}/2\pi$ & external optical coupling rate & 650 MHz \\     
    \hline
    $\kappa_\text{i,o}/2\pi$ & intrinsic optical loss rate & 650 MHz \\     
    \hline
    $\kappa_\text{m}/2\pi$ & acoustic loss rate & 150 kHz \\     
    \hline
    $\kappa_\text{e,mw}/2\pi$ & external coupling rate & 1.2 MHz \\     
    & of superconducting resonator & \\
    \hline
    $\kappa_\text{i,mw}/2\pi$ & intrinsic loss rate & 550 kHz \\
    & of superconducting resonator & \\
    \hline
    $\kappa_\text{mw}/2\pi$ & total damping rate & 1.75 MHz \\
        & of superconducting resonator & \\
    \hline
    \end{tabular}
    \caption{Frequencies and coupling rates of transducer internal modes.}
    \label{tab:params_coupling}
\end{table}

\begin{table}[!ht]
    \centering
    \begin{tabular}{|c|c|c|}
    \hline
    Symbol & Description   &  Value \\
    \hline
    $T_\text{p}$ & Two sigma duration of pump pulse  & 96 ns \\             
    \hline
    $T_\text{r}$ & Repetition period of experiment & 20 $\mathrm{\mu}$s \\             
    \hline
    $T_\text{d}$  &  Delay between pump pulses  & 279 ns \\   
    \hline
    - & Peak power of pump pulse & 83 nW \\             
    $n_a$ & Peak intra-cavity photon number & 0.78 \\    
    \hline
    $p_{\mathrm{click}} $ & Optical heralding probability & $5.2 \times 10^{-6}$\\
    & (single pulse) & \\
    \hline
    $R_{\mathrm{click}}$ & Optical heralding rate ($=p_{\mathrm{click}}/T_r$) & 0.26 s$^{-1}$ \\
    \hline
     $p$ & SPDC scattering probability & $1.0 \times 10^{-4}$\\
    \hline
    $\eta_{\mathrm{opt}}$ & Optical collection efficiency & $5.5 \times 10^{-2}$\\
    \hline
    $\eta_{\mathrm{mw}}$ & Microwave conversion efficiency & 0.59 \\
    \hline
    \end{tabular}
    \caption{Microwave-optical photon pair generation parameters.}
    \label{tab:params}
\end{table}

\begin{table}[!ht]
    \centering
    \begin{tabular}{|c|c|}
    \hline
    Parameter  &  Value \\
    \hline
     Optical cavity to on-chip waveguide  & 0.50 \\             
    \hline
    Waveguide to lensed fiber  & 0.27 \\             
    \hline
    Circulator between excitation and detection  & 0.90 \\             
    \hline
    \textbf{Total filter bank loss}  & \textbf{0.55} \\   
    \textbf{(individual components below)} & \\
    2x2 switches  & 0.79 \\    
    Filters (2x cascaded)  & 0.79  \\    
    Circulator in filter setup  & 0.88 \\        
    \hline
    SNSPD setup  & 0.83 \\   
    \hline
    \textbf{Optical collection efficiency ($\eta_{\mathrm{opt}}$)}  & $5.5 \times 10^{-2}$ \\   
    \hline
    \end{tabular}
    \caption{Independently calibrated losses of components along optical detection path for Z-basis measurements. In X-basis measurements, the time-delay interferometer setup adds extra transmission loss of 0.89.}
    \label{tab:params}
\end{table}

\section{Experimental setup}
\begin{figure*}
\includegraphics[width=\textwidth,scale=1]{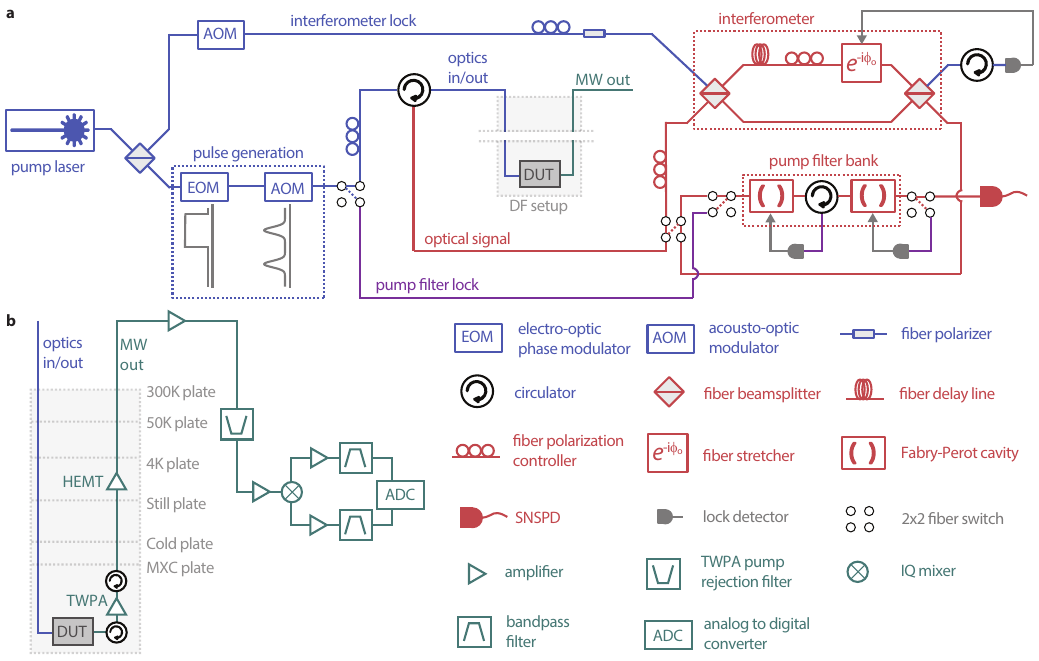}
\caption{\textbf{Schematic of experimental setup.} \textbf{a.} Schematic of optical setup highlighting essential components for the microwave-optical Bell state experiment. A more detailed schematic is provided in previous work \cite{Meesala2023}. Dotted rectangular blocks encompass the pump pulse generation module, the dilution fridge (DF) setup, interferometer used for X-basis optical measurements, and the pump filter bank. Individual components are labeled in the key below the schematic. \textbf{b.} Schematic of DF setup and the microwave amplification chain with Josephson traveling wave parametric amplifier (TWPA) mounted on the mixing chamber (MXC) plate and high electron mobility transistor (HEMT) amplifier mounted on the 4K plate.}
\label{si:optical_setup_fig}
\end{figure*}

The experimental setup used in this work is a modified version of the one used in our recent demonstration of non-classical microwave-optical photon pairs \cite{Meesala2023}. Fig.\,\ref{si:optical_setup_fig}a shows relevant upgrades to the optical setup that enabled us to generate and characterize microwave-optical Bell states. The microwave measurement chain shown in Fig.\,\ref{si:optical_setup_fig}b and the heralding setup are identical to the one used in Ref.\,\cite{Meesala2023}. Pump pulses used to excite the transducer are generated via amplitude and phase modulation of a continuous-wave external cavity diode laser. Our pulse generation setup uses two acousto-optic modulators (AOMs) in series for amplitude modulation and an electro-optic modulator (EOM) for phase modulation. By applying a rectangular voltage pulse of amplitude, $V_{\mathrm{eom}}$ to the phase modulator over the duration of the late pump pulse, we control the relative phase between the pump pulses, $\phi_\mathrm{p}$ introduced in the main text. Polarization of pump light sent to the device is controlled using an electronic fiber polarization controller (FPC). During long data acquisition runs, we mitigate the impact of long term polarization drifts along the excitation path through active polarization control. 

Pump pulses are sent to the transducer in the dilution fridge (DF) setup through a circulator, which allows us to direct optical emission along with reflected pump pulses to the detection setup. At the beginning of the detection path, a 2x2 switch allows us to route the signal to a home-built time-delay interferometer for X-basis measurements or to bypass it for Z-basis measurements. Prior to optical detection on a superconducting nanowire single photon detector (SNSPD), the signal is passed through a pump filter bank comprising two tunable Fabry-Perot cavities. These cavities are locked on resonance with the transducer optical resonance frequency, $\omega_{\text{o}}$ using a reference tone derived from the pump laser (`pump filter lock' path colored in purple in Fig.\,\ref{si:optical_setup_fig}a). Details of the locking procedure, which is performed once every four minutes during data acquisition, have been described previously \cite{Meesala2023}. For the current experiment, we upgraded to higher bandwidth filters (individual cavity bandwidth of 6MHz compared with 3.6MHz in the old setup), which allowed us to work with pump pulses of shorter duration. Including a circulator and two 2x2 fiber-optic switches, this upgraded filter bank has a total on-resonance insertion loss of 2.6dB at the signal frequency while providing 105dB extinction at the pump frequency.

Optical time-bin interference for X-basis measurements is realized in an asymmetric Mach-Zehnder interferometer placed inside a thermally insulated enclosure. A relative path difference of 53m is achieved via a fiber spool of fixed length in one of the arms of the interferometer. This path difference is pre-calculated and, upon insertion of the fiber spool into the setup, is experimentally verified to achieve the desired time delay, $T_\mathrm{d} = 279$ns between the optical time-bins. Optical emission from the transducer is directed into one of the two input ports of the interferometer while the single photon detection path is connected to one of the two output ports. The other input and output of the interferometer are used to transmit a lock tone (`interferometer lock' path colored in blue in Fig.\,\ref{si:optical_setup_fig}a), and actively stabilize the relative phase between the arms. Phase stabilization is achieved by measuring the power of the lock tone at the interferometer output on a photodetector, and controlling a piezoelectric fiber stretcher (Optiphase PZ1, 20kHz bandwidth) in the longer arm. PID feedback on the fiber stretcher is implemented using a Toptica Digilock 110 module. The interferometer lock tone is picked off from the pump laser via a beamsplitter upstream of the pulse generation module in the excitation path. Since the lock tone has the same center frequency as the pump pulses, it is largely extinguished by the pump filter bank on the single photon detection path. However, for the lock tone power of 400nW used in our experiment, we find significant bleedthrough of 500 counts/s on the SNSPD. To prevent false heralding events from this bleedthrough, an AOM is used to shut off the lock tone in a 2$\mathrm{\mu}$s window centered in time around the pump pulses used to excite transducer. This off-duration is much shorter than the drift timescale of the setup. A sample-and-hold setup on the electrical output of the lock photodetector is triggered in sync with the pulses applied to the lock tone. This ensures that the input to the PID controller does not drop to zero during the brief period for which the lock tone is turned off.

Before commencing X-basis measurements, polarization alignment for light incident on the output beamsplitter of the interferometer is performed by adjusting paddle FPCs shown in Fig.\,\ref{si:optical_setup_fig}a through the following procedure. First, a fiber polarizer inserted into the lock input arm allows us to define a fixed reference input polarization into the interferometer. The variable beamsplitters in the interferometer, which have a weak (few \%) polarization dependence, are balanced for this reference polarization in the lock input arm. Following this step, the FPC inside the interferometer enclosure is adjusted to maximize  interference visibility for the lock tone. Finally, polarization of light from the transducer arriving in the signal input arm is aligned to our fixed reference polarization in the lock arm. This is achieved by adjusting the FPC in the signal input arm to maximize the intensity of the RF beat note between signal and lock frequencies as observed on the lock photodetector. 

During X-basis data acquisition, the interference visibility for optical pulses emitted by the transducer is periodically monitored via a calibration procedure described in Section \ref{si:phase_cal_section}. We found that the passive stability of the interferometer setup was sufficient to maintain an average visibility of 94\% for signal pulses over the measurement duration of the X-basis dataset. Approximately 4\% reduction in visibility is due to finite overlap of the optical time-bins. The remaining 2\% reduction in visibility can be attributed to slow fluctuations of polarization along the signal collection path and of temperature inside the interferometer enclosure.

\section{Phase calibration for optical X-basis measurement}\label{si:phase_cal_section}

\begin{figure*}
\includegraphics[width=\textwidth,scale=1]{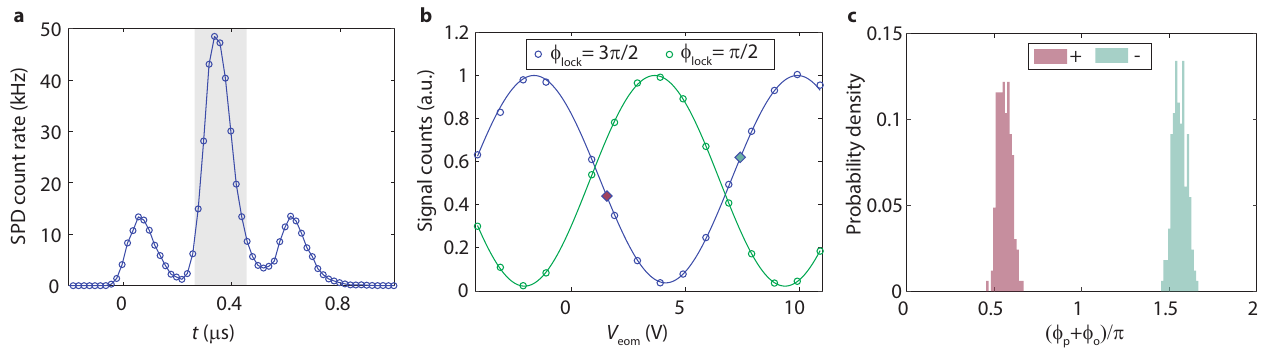}
\caption{\textbf{Optical phase calibration.} \textbf{a.} Time-binned histogram of optical detection events obtained from transmitting strong, coherent pulses generated by the transducer through the phase-stabilized interferometer setup. Gray shaded region indicates gating window used for interference detection. \textbf{b.} Typical phase calibration curve. Variation of gated optical counts in panel a with the EOM voltage, $V_{\mathrm{eom}}$ applied to the late pump pulse. Blue (green) data points correspond to calibration measurement with the interferometer locked to $\phi_{\mathrm{lock}} = 3\pi/2$ ($\pi/2$) at the frequency of the lock tone. Solid lines show cosine fits. Diamond markers indicate EOM voltages corresponding to the two phase settings, $\phi_\mathrm{p} + \phi_\mathrm{o}$ = $0.56\pi, 1.56\pi$ used for X-basis measurements of the Bell state. \textbf{c.} Histograms of calibrated phase, $\phi_\mathrm{p} + \phi_\mathrm{o}$ between the optical time-bins for the two phase settings used in the X-basis measurement (denoted by $+$ and $-$ in the legend). The histograms are generated from 268 calibration routines interleaved with data acquisition over the course of the measurement.}
\label{si:optical_phase_cal}
\end{figure*}

During X-basis measurements, the fiber stretcher in Fig.\,\ref{si:optical_setup_fig}a is actively stabilized to lock the interferometer at the halfway point on interference fringes generated by the lock tone on the output detector. This corresponds to a relative phase, $\phi_{\mathrm{lock}} = 3\pi/2$ between the interferometer arms at the lock tone frequency. Since the optical signal from the transducer has a different center frequency than the lock tone, a different but fixed relative phase, $\phi_\mathrm{o}$ is imprinted between the two time-bins in the signal. To calibrate $\phi_\mathrm{o}$, the transducer is optically pumped with the double Gaussian pulse sequence used in the entanglement experiment. At the same time, the  microwave port of the transducer is excited with a CW electrical tone at the microwave resonance frequency, $\omega_{\text{m}}$. Under simultaneous optical and microwave excitation in this manner, the transducer generates two large-amplitude coherent states in the optical early and late modes expected in the SPDC signal. After passing these coherent optical pulses through the phase-stabilized interferometer and the pump filter bank, we measure a time trace of the resulting optical intensity on the SNSPD by histogramming photon arrival times as shown in Fig.\,\ref{si:optical_phase_cal}a. Fig.\,\ref{si:optical_phase_cal}b shows a typical phase calibration curve where gated counts in the interference window are measured as the phase between the pump pulses, $\phi_\mathrm{p}$ is swept by controlling $V_{\mathrm{eom}}$, the amplitude of the voltage pulse applied to the EOM in the pulse generation module in Fig.\,\ref{si:optical_setup_fig}a. A cosine fit to such an interference curve with the expression, $A\mathrm{cos}(kV_{\mathrm{eom}} + \phi_{\mathrm{o}}) + B$ allows us to calibrate both $\phi_\mathrm{p} =kV_{\mathrm{eom}} $ and $\phi_\mathrm{o}$. Sweeping $\phi_\mathrm{p}$ via $V_{\mathrm{eom}}$ produces an effect equivalent to sweeping $\phi_\mathrm{o}$ via the interferometer lock setpoint, $\phi_\mathrm{lock}$. This is shown by the two interference curves in Fig.\,\ref{si:optical_phase_cal}b which are taken for two lock settings of the interferometer where $\phi_\mathrm{lock}$, and hence, $\phi_\mathrm{o}$ is changed by $\pi$. From an experimental standpoint, controlling $\phi_\mathrm{p}$ and fixing $\phi_\mathrm{o}$ ensures identical locking stability for the interferometer for the optical measurement phases of interest in our X-basis measurements. During a long X-basis data acquisition run, we periodically verify our calibration for the optical phase by performing this routine every two hours. Fig.\,\ref{si:optical_phase_cal}c shows a histogram of the total optical phase, $\phi_\mathrm{p} + \phi_\mathrm{o}$ from calibration runs performed over the month-long duration of the measurement, indicating nearly normal distributions with standard deviation of $0.04\pi$. 
% In addition, small offsets in the center frequency and polarization of the calibration pulses relative to the single photon pulses can produce a systematic error in the estimate of $\phi_\mathrm{p} + \phi_\mathrm{o}$. 

\section{Microwave mode envelope}\label{sec:MW_envelope}
In our experiments, the superconducting electrical resonator is tuned into resonance with the transducer acoustic mode such that the electro-acoustic coupled mode system is at its maximal hybridization point. The piezoelectric coupling rate, $g_\text{pe}$ and the damping rates of both modes are determined via microwave reflection spectroscopy of the device \cite{Meesala2023} and tabulated in Table \ref{tab:params_coupling}. The complex eigenvalues of the electro-acoustic coupled mode system are given by \cite{Zhong2020FreqBin},
\begin{equation}
\lambda_\pm = i\omega_\text{m}-\frac{\kappa_\text{m}+\kappa_\text{mw}}{4} \pm \sqrt{\left(\frac{\kappa_\text{m}-\kappa_\text{mw}}{4} \right)^2-g_\text{pe}^2}
\end{equation}
A single phonon generated by SPDC at $t=0$ is emitted with efficiency, $\eta_{\text{mw}} = 0.59$ into a temporal mode in the microwave waveguide defined by the envelope function,
\begin{equation}
f(t) = c \Theta(t) \left( e^{\lambda_+ t} - e^{\lambda_- t}  \right),    
\end{equation}
where $\Theta(t)$ is a Heaviside step function and $c$ is a normalization constant satisfying $ \int|f(t)|^2 dt = 1$. In our device, since  $2g_\text{pe} > \kappa_\text{mw}, \kappa_\text{m}$, the electrical and acoustic resonators are strongly coupled and the exponentially decaying envelope $f(t)$ exhibits oscillations with a period, $2\pi/|\lambda_+ - \lambda_-|$. When two optical pump pulses excite the device at $t=0, T_\text{d}$, the corresponding microwave emission envelopes, $f(t), f(t-T_{\text{d}})$ are exactly orthogonal when $T_\text{d} = \pi/|\lambda_+ - \lambda_-| = 279$ns. In Fig.\,1c of the main text, we plot $|f(t)|^2$ and $|f(t-T_\text{d})|^2$ as the intensity envelopes of the theoretically expected early and late microwave modes. For Z-basis intensity measurements, we convolve the digitized, down-converted microwave  voltage signal, $V(t)$ with $f$ and estimate the microwave quanta, $|\int f^{*}(t-\tau)V(t)dt|^2/(2{Z_\mathrm{o}}G\hbar\omega_\mathrm{m})$, where $\tau$ is a variable readout delay, $Z_\mathrm{o} = 50\Omega$, and $G=107.4$dB, the gain of the amplification chain referred to the output of the transducer chip, is calibrated via optomechanical sideband asymmetry and microwave spectroscopy \cite{Meesala2023}. After subtracting independently measured amplifier-added noise in the same mode as described in Sec.\,\ref{sec:mw_data_proc}, the microwave output quanta from the transducer are plotted in Fig.\,2c of the main text with varying $\tau$. We determine optimal readout delays, $T_{\text{e}}$ and $T_{\text{l}}=T_{\text{e}}+T_{\text{d}}$, which maximize the microwave-optical cross-correlation function, and define microwave early and late modes with the envelope functions, $f(t-T_\text{e}), f(t-T_\text{l})$, respectively. 

The complex-valued voltages for the Z-basis microwave modes detected at the output of the heterodyne setup are given by
\begin{subequations}
\begin{align}
 S_{\text{e}} &= \int f^{*}(t-T_\text{e})V(t) dt \\
 S_{\text{l}} &= \int f^{*}(t-T_\text{l})V(t) dt
\end{align}
\end{subequations}
For the X-basis microwave modes, the complex voltage is given by $S_{\text{e}} + e^{i\phi_{\text{m}}}S_{\text{l}}$ where the measurement phase, $\phi_{\text{m}}$ can be chosen in post-processing.

\section{Microwave data processing}\label{sec:mw_data_proc}
Our microwave data processing approach follows techniques originally developed in Ref.\,\cite{daSilva2010,Eichler2011} for quantum state tomography of itinerant microwave photons. These techniques have been used in recent demonstrations of entangled microwave photons \cite{Kannan2020, Besse2020, Ferreira2022}. With the goal of subtracting amplifier-added noise, modes at the output of the heterodyne setup are related to the modes exiting the transducer via the input-output relation of a phase insensitive bosonic amplifier\cite{Caves1982,daSilva2010},
\begin{subequations}\label{eq:bos_amp}
\begin{align}
\hat{S}_\text{e} &= \sqrt{G}\hat{C}_\text{e} + \sqrt{G-1} \hat{H}^\dagger_\text{e}, \\
\hat{S}_\text{l} &= \sqrt{G}\hat{C}_\text{l} + \sqrt{G-1}\hat{H}^\dagger_\text{l}.
\end{align}
\end{subequations}
Here $G$ is the gain of the amplification chain referred to the output of the transducer chip. $\hat{H}_\text{e}$ $(\hat{H}_\text{l})$ represent amplifier noise added to the early (late) modes respectively. We define complex tensors, $\bar{C}$ and $\bar{S}$ containing the normal ordered moments of the microwave modes at the output of the device and the output of the heterodyne setup, respectively, given by
\begin{align}
\bar{C}_{wxyz} &:= \langle \hat{C}_\text{e}^{\dagger w} \hat{C}_\text{e}^{x} \hat{C}_\text{l}^{\dagger y} \hat{C}_\text{l}^{z} \rangle,  \label{eq:C_bar}\\
    \bar{S}_{klmn} &:= \langle \hat{S}_\text{e}^{\dagger k} \hat{S}_\text{e}^{l} \hat{S}_\text{l}^{\dagger m} \hat{S}_\text{l}^{n} \rangle\label{eq:S_bar}.
\end{align}
We also define the tensor $\bar{H}$ containing the anti-normal ordered moments of the amplifier noise,
\begin{equation} \label{eq:H_bar}
     \bar{H}_{klmn} := \langle \hat{H}_\text{e}^k \hat{H}_\text{e}^{\dagger l} \hat{H}_\text{l}^m \hat{H}_\text{l}^{\dagger n} \rangle,
\end{equation}
which is expected to be a product of  thermal states in the early and late modes.

Using Eqs.\ref{eq:bos_amp}-\ref{eq:H_bar} in the limit $G\gg1$ and adopting the multi-index notation, $\alpha=(k,l,m,n)$ and $\beta = (w,x,y,z)$, we relate the moment tensors through the multinomial expansion,
\begin{equation}\label{eq:moment_expansion}
    \bar{S}_\alpha =  \sum_{\beta\leq \alpha}\bar{T}_{\alpha\beta} \bar{C}_\beta,
\end{equation}
where,
\begin{equation}
   \bar{T}_{\alpha\beta} = G^{|\alpha|/2} \binom{\alpha}{\beta}\bar{H}_{\alpha-\beta}.
\end{equation}

As described in Sec.\,\ref{sec:MW_envelope}, we apply a matched filter to the heterodyne data in our experiments and obtain a dataset of complex voltage samples $\{S_\text{e},S_\text{l}\}$ for each optical detection outcome. We estimate the elements of the moment tensor, $\bar{S}$ by taking the sample mean,
\begin{equation}
\bar{S}_{\alpha} =\frac{1}{N} \sum  (S^{*}_{\text{e}})^k S^l_{\text{e}} (S^{*}_{\text{l}})^m S^n_{\text{l}}, 
\end{equation}
where $N$ is the number of voltage records in a given dataset. We interleave every fifteen minutes of conditional microwave data acquisition with calibration measurements where the transducer is (i) not pumped optically, and (ii) excited on the microwave port with a weak coherent state of fixed amplitude. (i) yields voltage samples of the amplifier noise and is used to construct $\bar{H}$, (ii) is used to track small fluctuations in the amplifier noise and gain which occur over the timescale of several hours. With this knowledge, we invert the moments at the output of the heterodyne setup, $\bar{S}_\alpha$ via the linear relation in Eq.\,\ref{eq:moment_expansion} to recover the moment tensor at the device output, $\bar{C}_\beta$. We estimate the statistical uncertainty, $\sigma_\beta$ on each moment using the sample variance. We then reconstruct the density matrix of the conditional microwave state, ${\rho}^{{(i)}}$ prepared by a single photon detection event in the optical mode $i \in \{\text{e,l},+,-\}$.

For each optical outcome $i$, the conditional density matrix is obtained by minimizing the log likelihood function, 
\begin{equation}
-\log\mathcal{L}({\rho}^{(i)}) = \sum_{\beta} \frac{|\bar{C}_{\beta}-P_{\beta}|^2}{\sigma_{\beta}^2}.
\end{equation}
Here $P_{\beta}=\text{Tr}\left\{\hat{C}^{\dagger w}_\text{e} \hat{C}_\text{e}^x \hat{C}^{\dagger y}_\text{l} \hat{C}_\text{l}^z {\rho^{(i)}} \right\}$ is the expectation value for the moment, $\bar{C}_j$ predicted by the underlying density matrix. We solve the constrained optimization problem,
\begin{align} \label{eq:logmin}
    \min_{\rho} \;\;\; & -\log\mathcal{L}(\rho^{(i)}) \\
    \text{s.t.}\;\;\; & \text{Tr}\{\rho^{(i)}\} = 1 \nonumber\\ 
     \;\;\;\; & \rho^{(i)} \succeq 0 \nonumber
\end{align}
using CVX, a package for solving convex programs \cite{cvx2012,grant2008}. The results for the maximum likelihood (ML) estimates for ${\rho}^{(i)}$ shown in the main text are obtained from a Z-basis (X-basis) dataset consisting of microwave voltage records associated with $\approx 3\times10^5$ ($\approx 7\times10^4$) heralding events in each of the early and late ($+$ and $-$) optical modes acquired over one month. We reconstruct each ${\rho}^{(i)}$ in a truncated 6x6 Fock space of early and late modes which we found to be sufficient to achieve convergence in the single photon probabilities. In Fig.\,4a of the main text, we plot ${\rho}^{(i)}$ in a truncated space of up to two photons in each mode for easy visualization. 

\begin{figure}[ht]
\includegraphics[width=0.48\textwidth]{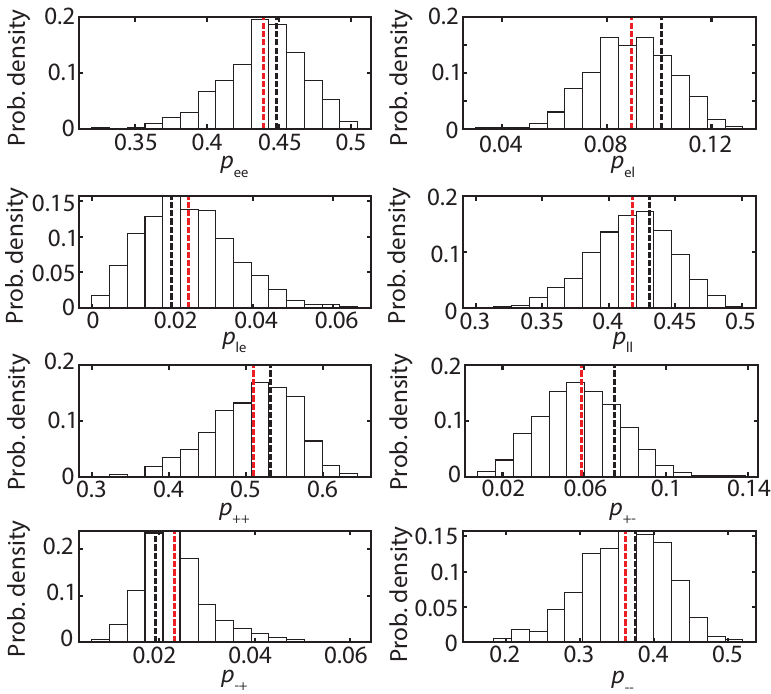}
\caption{\textbf{Error analysis for conditional  probabilities.} Marginal distributions for each conditional single microwave photon probability calculated by parametric bootstrapping and displayed as a histogram. The mean of the distribution is indicated by the red dashed line, and the result from maximum likelihood estimation over the datasets is shown by the black dashed line.}
\label{si:error}
\end{figure}

To determine the statistical uncertainty on the ML density matrices, we employ a parametric bootstrapping procedure \cite{Chow2012} with $10^3$ re-sampling iterations over the datasets with conditional heterodyne voltage samples. In each bootstrapping iteration, a sample of $\rho^{(i)}$ is generated by executing the optimization routine in Eq.\,\ref{eq:logmin} with moments randomly sampled from a multivariate normal distribution with the experimentally measured mean $\bar{C}_\beta$ and variance $\sigma_\beta^2$. From the ensemble of conditional density matrices generated with this procedure, we estimate marginal distributions associated with the conditional  probabilities, $p_{ij}$ of a single microwave photon in mode $j$ given a single optical photon detection event in mode $i$ as defined in Section \ref{sec:fidelity}. Here $i,j\in \{\text{e,l}\}$ for Z-basis measurements and $\{+,-\}$ for X-basis measurements. These marginal distributions are shown in Fig.\,\ref{si:error}. It is well-established that there can be slight bias between the mean of the distribution generated from bootstrapping and the ML estimate of the density matrix due to a combination of nonlinearities in the estimator and physical constraints on the density matrix \cite{Keith2018}. In our case, we calculate a mean value for the Bell state fidelity lower bound, $F_\text{lb}=0.83$ using the mean values of the conditional probabilities from the bootstrapping procedure (red vertical lines in Fig.\,\ref{si:error}), which is slightly higher than value of 0.79 from the ML estimates for the same probabilities (black vertical lines in Fig.\,\ref{si:error}). To take such systematic bias into account, we generate conservative confidence intervals for each $p_{ij}$ as $[ \text{min}\{\mu_{ij},p_{ij}^{\text{ML}}\}-\sigma_{ij}, \text{max}\{\mu_{ij},p_{ij}^{\text{ML}}\}+\sigma_{ij}]$, where $\mu_{ij}, \sigma_{ij}$ are the mean and standard deviation of the distributions, respectively and $p_{ij}^{\text{ML}}$ is the ML estimate from the dataset. Using these conservative confidence intervals, shown with the error bars on Fig.\,4b of the main text, we estimate $F_\text{lb} = 0.794^{+0.048}_{-0.071}$ which is above 0.5 by over four standard deviations.

\section{Bell state fidelity lower bound}\label{sec:fidelity}
To obtain the probability of a single photon in microwave mode $j$, we define projectors $\{ \Pi^{(j)} \}$ in the microwave subspace,
\begin{subequations}
\begin{align}
\Pi^{\text{(e)}} &= \ket{10}\!\bra{10}, \\
\Pi^{\text{(l)}} &= \ket{01}\!\bra{01}, \\
\Pi^{{(+)}} &= \frac{1}{2} \left(\ket{10} + e^{-i\phi_\text{m}}\ket{01} \right)\!\left(\bra{10} + e^{i\phi_\text{m}}\bra{01}\right), \\
\Pi^{{(-)}} &= \frac{1}{2}\left(\ket{10} - e^{-i\phi_\text{m}}\ket{01} \right)\!\left(\bra{10} - e^{i\phi_\text{m}}\bra{01}\right).
\end{align}   
\label{eq:projectors}
\end{subequations}
The conditional single-photon probabilities, $p_{ij}$ associated with a single optical photon detection event in mode $i$ are then calculated as 
\begin{equation}
    p_{ij} = \frac{\text{Tr} \left\{ \Pi^{(j)}\rho^{(i)} \right\}}{\Sigma_{i,j}\text{Tr} \left\{ \Pi^{(j)}\rho^{(i)} \right\}},
    \label{eq:cond_prob}
\end{equation}

\noindent where $i,j\in \{\text{e,l}\}$ for Z-basis measurements and $\{+,-\}$ for X-basis measurements. The normalization term in the denominator corresponds to post-selection of the single microwave photon subspace. These conditional single photon probabilities enable us to bound the Bell state fidelity using Eq.\,(2) of the main text \cite{Zhong2020TimeBin, Blinov2004}. 

\section{Contribution of two photon events to the heralding signal}
Our analysis for the Bell state fidelity uses the condition that the optical heralding events used for conditional microwave readout are due to detection of single photons. In our experiments, the fraction of optical heralding events due to terms with more than one optical photon is expected to be of the order of $p=1\times10^{-4}$, a negligible fraction of the dataset. These events do not have a statistically significant impact on the lower bound for the Bell state fidelity.

\section{Model for the entangled state}

\begin{table}[!ht]
    \centering
    \begin{tabular}{|c|c|c|}
    \hline
    Parameter  &  Value \\
    \hline
    $\eta_{\text{ext}}$ & 0.42\\
    \hline
    $\bar{n}_{\text{i,e}}$ & 0.05/$\eta_{\text{ext}}$\\
    \hline
    $\bar{n}_{\text{i,l}}$ & 0.10/$\eta_{\text{ext}}$\\
    \hline
    $\bar{n}_{\text{d,e}}$ & 0.029/(1-$\eta_{\text{ext}}$)\\
    \hline
    $\bar{n}_{\text{d,l}}$ & 0.029/(1-$\eta_{\text{ext}}$)\\
    \hline
    \end{tabular}
    \caption{Parameters for the conditional microwave state model. Thermal noise parameters are independently determined from the unconditional microwave signal shown by the dashed time trace in Fig.\,2c of the main text. Microwave extraction efficiency, $\eta_{\text{ext}}$ is determined from an independent measurement of the conditional microwave intensity when the transducer is excited with a single pump pulse. $\eta_{\text{ext}}$ is less than the theoretical microwave conversion efficiency, $\eta_{\text{mw}}$=0.59 due to imperfections in mode-matching with the matched filter.}
    \label{tab:params_model}
\end{table}

\begin{figure}[ht]
\includegraphics[width=0.48\textwidth]{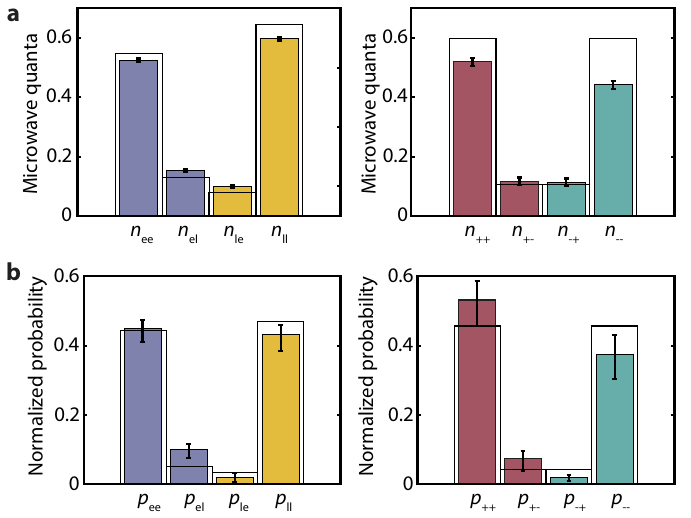}
\caption{\textbf{Model for the entangled state.} In both panels, predictions from the model are given by the white bars while measured values and associated uncertainty are given by the colored bars. \textbf{a.} Conditional microwave intensity, $n_{ij}$ in the microwave mode $j$ conditioned on receipt of an optical click in mode $i$. \textbf{b.} Conditional probability, $p_{ij}$ of a single photon in microwave mode $j$ conditioned on receipt of an optical click in mode $i$. }
\label{si:model}
\end{figure}

Here we present a numerical model for the microwave-optical entangled state by incorporating the level of pump-induced noise in the transducer. This is an extension of a simple model presented in our previous work on microwave-optical photon-pair generation \cite{Meesala2023}. The conditional microwave state, $\rho^{(i)}$ corresponding to a single optical photon detection event in mode $\hat{A}_i$, where $i \in \{\text{e,l},+,-\}$ is generated in two steps from an initial thermal state, $\rho_{\text{th}}$ with occupation $\bar{n}_{\text{i,e}}$ ($\bar{n}_{\text{i,l}}$) in the early (late) microwave mode prior to the corresponding pump pulse. 
\begin{enumerate}
    \item Detection of a single optical photon in mode $\hat{A}_i$ signifies successful pair generation and conditionally adds a single excitation to the microwave mode $\hat{C}_i$. The microwave state undergoes a quantum jump into a photon-added thermal state: $\rho_{\text{th}} \rightarrow \hat{C}_i^\dagger \rho_{\text{th}} \hat{C}_i/\text{Tr}\{\hat{C}_i^\dagger \rho_{\text{th}} \hat{C}_i\}$. 
    \item This photon-added thermal state is extracted with finite efficiency, $\eta_{\text{ext}}$ and additional thermal noise quanta, $\bar{n}_{\text{d,e}}$ ($\bar{n}_{\text{d,l}}$) added to the early (late) microwave mode due to delayed heating of the transducer after the pump pulse. These effects are together modeled as an imperfect beamsplitter transformation on both early and late modes. We have $\hat{C}_e \rightarrow \sqrt{\eta_{\text{ext}}}\hat{C}_e + \sqrt{1-\eta_{\text{ext}}}\hat{d}_\text{e}$ where $\hat{d}_\text{e}$ is a phenomenological thermal noise operator such that $(1-\eta_{\text{ext}})\langle \hat{d}_\text{e}^\dagger \hat{d}_\text{e} \rangle = \bar{n}_{\text{d,e}}$, and similarly for the late mode. Upon applying unitary operators for these beamsplitter transformations on the photon-added thermal state, we obtain the conditional microwave state, $\rho^{(i)}$.
\end{enumerate}

The occupation of microwave mode $j$ correlated with a single photon click in optical mode $i$ is then given by $n_{ij} = \text{Tr}\{\hat{C}_j^\dagger \hat{C}_j \rho^{(i)}\}$. Conditional single photon probabilities, $p_{ij}$ are calculated from the modeled $\rho^{(i)}$ using equations \ref{eq:projectors},\ref{eq:cond_prob}. The model is implemented in QuTiP \cite{Johansson2013} in the joint Fock space of early and late microwave modes truncated to a Fock size of 6 for each mode, which was determined to be sufficient for convergence of  $n_{ij}$ and $p_{ij}$. We use noise parameters tabulated in Table \ref{tab:params_model} determined from independent measurements. The results from the model plotted with white rectangles in Fig.\,\ref{si:model} allow us to predict visibilities, $V_z = V_x = 0.70$ for the conditional microwave intensities, and a fidelity lower bound, $F_{\text{lb}} = 0.83$ after post-selecting the single microwave photon subspace. These predicted values from the model are slightly higher than the experimentally measured values which are impacted by dark counts, and imperfections in optical time-bin interference.

\bibliography{refs_SI}